\documentclass[%
 reprint,
 amsmath,amssymb,
 aps,
]{revtex4-1}

\usepackage{graphicx}
\usepackage{dcolumn}
\usepackage{bm}
\usepackage{siunitx} 
\usepackage[caption=false]{subfig}

\begin{document}


\title{Detection of Applied and Ambient Forces with a Matterwave Magnetic-Gradiometer}

\author{Billy I. Robertson}
\author{Andrew R. MacKellar}
\author{James Halket}
\author{Anna Gribbon}
\author{Jonathan D. Pritchard}
\author{Aidan S. Arnold}
\author{Erling Riis}
\author{Paul F. Griffin}

\affiliation{Department of Physics, SUPA, University of Strathclyde, Glasgow, G4 0NG, United Kingdom\\}

\date{\today}

\begin{abstract}
An atom interferometer using a Bose-Einstein condensate of $^{87}$Rb atoms is utilized for the measurement of magnetic field gradients. Composite optical pulses are used to construct a spatially-symmetric Mach-Zehnder geometry. Using a biased interferometer we demonstrate the ability to measure small residual forces in our system and discriminate between magnetic and intertial effects.. These are a residual ambient magnetic field gradient of 15$\pm$2~mG/cm and an inertial acceleration of 0.08$\pm$0.02~m/s$^{2}$. Our method has important applications in the calibration of precision measurement devices and the reduction of systematic errors.


\begin{description}
\item[PACS numbers] 03.75.Dg, 39.20.+q, 03.75.Be.
\end{description}
\end{abstract}

\pacs{Valid PACS appear here}
\maketitle

\section{\label{sec:intro}Introduction}

Matter-wave interferometry is performed by splitting the atomic wavefunction into two or more distinct parts, then allowing these to evolve before recombining them and performing a readout operation \cite{Pritchard09}. Alkali atoms offer an attractive medium due to the relatively simple internal structure, and the ability to readily prepare, manipulate, and interrogate them with lasers \cite{Pritchard09,Kasevich00}. Atom interferometry in general offers the ability to perform precision measurements in the form of magnetic and gravitational metrology \cite{Kasevich98,Wu05,Luo10,Kumarakrishnan11,Biedermann14,Tino15,Rasel16,Robins16} and inertial sensing \cite{Robins16,Borde91,Kasevich97,Pritchard97,Kasevich00,Kasevich06,Rasel15} as well as to test fundamental physics by making measurements of the fine-structure constant \cite{Pritchard02,Chu02,Biraben06,Biraben11}, the Newtonian gravitational constant \cite{Tino08}, atomic polarisabilities \cite{Sackett08}, tests of the equivalence principle \cite{Weitz04}, and recent proposals for gravitational wave detection \cite{Rajendran13}. The use of Bose-Einstein condensates over thermal atoms in interferometry can be favourable due to the low atomic speed and therefore low dispersion, and increased phase coherence offering a high--contrast signal and an increased signal-to-noise ratio \cite{Robins11,Rasel16,Hagley00}.

In this paper we present an application of atom interferometry to the measurement of magnetic and inertial field gradients. In the first instance we use measurements of the varying momentum-state populations at the interferometer output to measure an applied magnetic field gradient along the interferometer axis  \cite{Robertson_Thesis}. Our apparatus is operated in two modes using either pulsed and continuous magnetic field gradients. We model the corresponding interferometer phases and demonstrate the effect of varying the magnitude of the gradient field as well as varying the duration of the interferometer sequence.

Through the application of bias magnetic field gradients we further demonstrate the ability to clearly distinguish the effects of both applied and ambient magnetic and ambient inertial gradient fields. This result enables the characterization of otherwise small residual accelerating fields, both magnetic and inertial, within the complete system that would otherwise be unresolvable in an interferometer with a cosine-dependent phase sensitivity.

\section{\label{sec:prep}Preparation of BEC}

We load a 3D magneto-optical trap (MOT) of $\approx10^9$  $^{87}$Rb atoms from a 2D MOT in 15~s. After optical molasses and optical pumping into the $\left|F=2,\rm{m}_F=2\right\rangle$ ground state we magnetically transport our atoms horizontally to the interferometry region, 6~cm from the 3D MOT, and compress the cloud by ramping the quadrupole (QP) trap to an axial gradient of 206~G/cm. Starting with an initial phase-space density of 10$^{-6}$, we perform radio frequency (RF) evaporative cooling by exponentially ramping an RF field from 16~MHz to 3.4~MHz over 3~s. After this stage $5\times10^7$ atoms remain with a temperature of 32~$\mu$K. We then adiabatically transfer the atoms into a crossed optical dipole trap over 200~ms whilst the RF field is ramped exponentially from 3.4~MHz to 0.6~MHz and the QP field is ramped linearly from 206~G/cm to 15~G/cm, where the final value counteracts the acceleration due to gravity \cite{Porto09}. Our crossed optical dipole trap comprises two intersecting 1070~nm beams, each of 2.5~W and focused to a waist of $w_0=86~\mu$m positioned $\sim100~\mu$m below the QP centre. At this point we have an optical-magnetic hybrid trap containing $\approx4\times10^6$ atoms.

We continue the forced evaporative cooling by exponentially ramping the power of the dipole beams from 2.5~W to 300~mW over 4~s, and  ramping the QP field from 15~G/cm to 7.5~G/cm over the first 100~ms~\cite{Porto09}. Halfway through this stage we apply a large (20~G) bias field in the vertical direction to move the QP centre up by $\approx$2.7~cm; this allows for a magnetic gradient with reduced curvature for use as a levitation field, in addition to a magnetic launch during a later stage of the experiment. At the end of the power ramp we hold the power constant for $\approx$300~ms. We now have a BEC of $\approx1\times10^5$ atoms and $>$80~\% purity at a temperature of $\approx$100~nK in the $|F=2,m_{\rm{F}}=2\rangle$ state with which we can perform interferometry. Further details on the creation of our BEC can be found in~\cite{Robertson_Thesis}.

\section{\label{sec:sequence}Interferometer Sequence}

\begin{figure}
	\centering
	\includegraphics[width=1\linewidth]{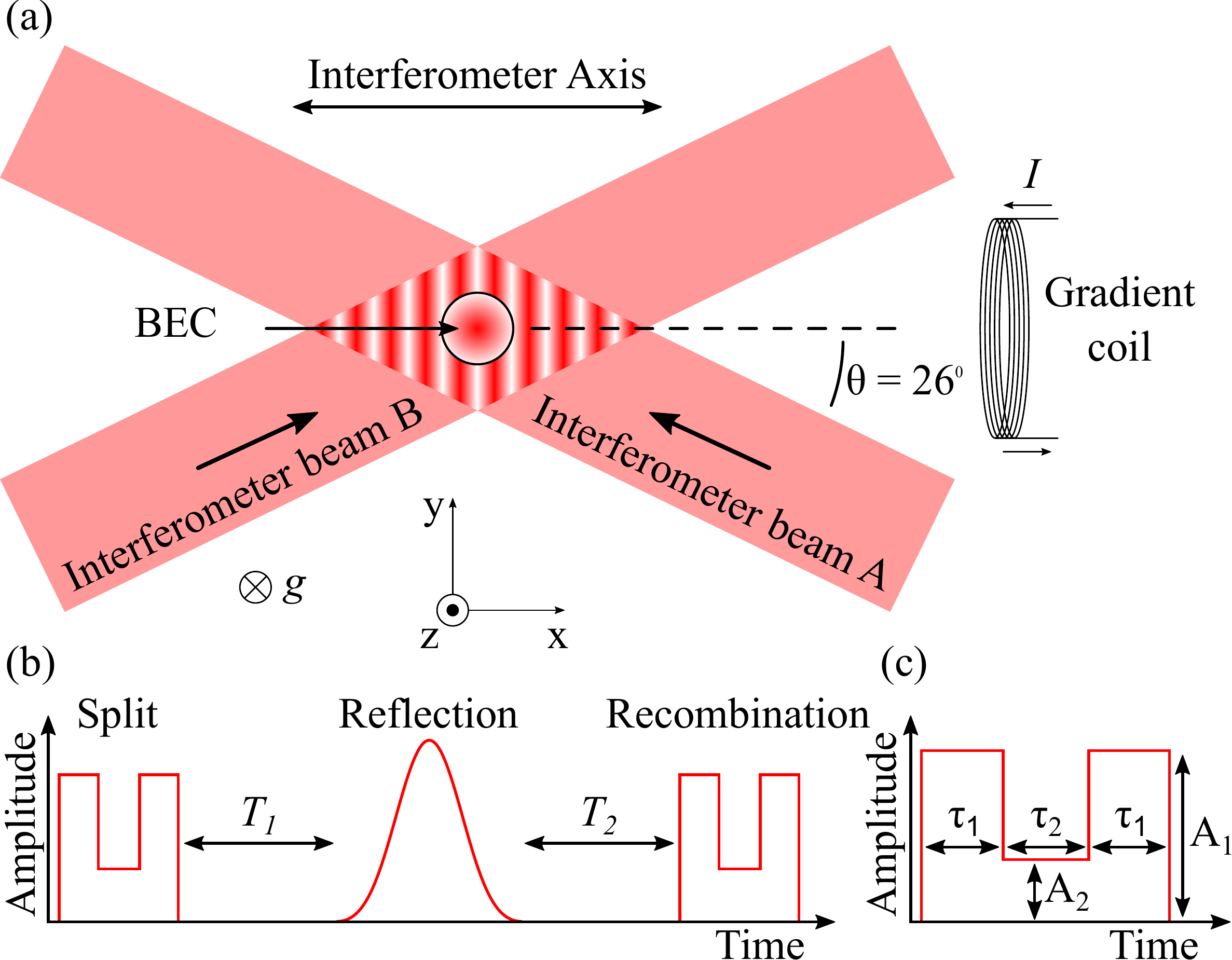}
	\caption{(a) The interferometer beams of wavelength 780~nm cross at a half--angle of $\theta$=26\si{\degree} creating an optical lattice of period 434~nm along the x-axis. We define this as our interferometer axis. The magnetic gradient coil is positioned in the $y-z$ plane and on-axis with the interferometer. (b) Interferometer pulse sequence. (c). Breakdown of composite splitting and recombination pulses.}
	\label{fig:experiment_diagram}
\end{figure}

We operate our matterwave interferometer using Kapitza-Dirac-type beam--splitter pulses, operating in the so-called Raman-Nath regime~\cite{Pritchard09, Zeilinger95, Gupta01}, populating atoms symmetrically in both the negative and positive momentum modes of $|p=\pm2\hbar k\rangle$. In contrast to the widely-used resonant-Bragg scheme, in this regime the optical pulses are provided by two laser beams of identical frequency and, hence, they provide off-resonant coupling between two-photon-coupled momentum states. However, a result of the detuned interaction is that 100\% transfer to a target momentum state is not possible by single square pulses. To circumvent this, multi--pulse techniques have been proposed \cite{Prentiss05,Clark10} 
and implemented \cite{Wu05} that show near-perfect fidelity of the beam-splitter process. These works made use of two short Kapitza-Dirac optical grating pulses, separated by a free--evolution period to allow the relative phase of the atomic momentum modes to shift by $\pi$, creating destructive interference in the stationary, $|p=0\rangle$ mode. In our work we optimize the pulse parameters by numerical modelling of Raman-Nath equations \cite{Prentiss05}, with additional control over the coupling Rabi frequencies during the rephasing stage \cite{Griffin17}.

We perform interferometry by the interaction of the BEC with two laser beams of wavelength 780~nm, crossing at a half--angle of 26\si{\degree} as shown in Fig. \ref{fig:experiment_diagram}, and focussed to a waist ($e^{-2}$ radius) of $w_0\approx95~\mu$m. The laser is locked to the $^{85}$Rb $\left|F=1\right\rangle\rightarrow\left|F'=1,2\right\rangle$ crossover transition for convenience, which is $\approx$4~GHz blue detuned from the closest optical resonance of our BEC. The laser intensity is controlled via an AOM, to which we send arbitrary pulse sequences via an SRS DS345 arbitrary function generator. The beam is then split into two beams, A and B, and overlapped with the two arms of the crossed dipole trap as shown in Fig.~1(a). To ensure symmetric splitting, these beams are mode-matched to locate the beam waist at the intersection. Both interferometer beams are vertically polarized and, as such, they drive the linear atomic Raman transitions and the atoms remain in the same internal state after each interferometer pulse. This ensures that all atoms contribute to the interferometer, and avoids the blow-away pulses required in other interferometer schemes to remove untargeted states (eg.~\cite{Rasel15,Biraben06}). Additionally, since the atoms in both arms are in the same internal state, the interferometer is insensitive to spatially-uniform magnetic fields.

After we create our BEC we release it from the hybrid trap and allow 6~ms of mean-field expansion to reduce self-interactions \cite{Gupta11}. Gravitational acceleration during this expansion would give a downward velocity of 59~mm/s, causing the atoms to fall out of the interferometer beams. We therefore apply a magnetic launch sequence during this time in which we ramp the levitating magnetic field from 15~G/cm to 21.8~G/cm in the first 4~ms, followed by an exponential switch-off over 2~ms. This weak positive acceleration ensures atoms start above the beam centre with a low downward velocity of 2~mm/s, maximizing the interferometer interrogation time. At the end of the 6~ms expansion the only field remaining is a bias field of 1.02~G in the z-axis, used to maintain a quantisation axis. 

The interferometer sequence is shown in Fig. \ref{fig:experiment_diagram}b. Both interferometer beams have the same frequency and therefore produce a static optical lattice a static optical lattice with lattice vector $k=2\pi\cos(\theta)/\lambda$.
We characterize the amplitude of the optical pulses in terms of recoil energy $E_{\rm{r}} = \hbar^2 k^2/\left(2m\right)$, where $\hbar$ is the reduced Planck constant, $m$ is the mass of the atom. 
By using this metric the pulse parameters are easily transferable between atom species. We first apply a composite splitting pulse (see Fig. \ref{fig:experiment_diagram}c) of $\tau_1=26.6~\mu$s, $\tau_2=45.6~\mu$s, $A_1=6.07~E_{\rm{r}}$, and $A_2=0.52~E_{\rm{r}}$,  which separates the wavefunction into the $\pm2\hbar k$ momentum states with almost 100\% efficiency, where the pulse parameters are optimized, as outlined earlier. After some time, $T_1$, we reverse the momenta of the wavepackets by a mirror pulse with a continuous temporal profile. A Blackman pulse, with intensity profile
\begin{equation}
	\resizebox{.9\hsize}{!}{$ 
	y(t) = A\left(0.427 - 0.497\,\cos{\left(\frac{2\pi (t-t_0)}{\tau}\right)}+0.077\cos{\left(\frac{4\pi (t-t_0)}{\tau}\right)}\right)~,
	$}
\end{equation} 
for $t_0 \le t \le t_0+\tau$, is used for the mirror sequence, as this shape has supressed side lobes in the frequency spectrum when compared to the square pulses used in the beam--splitter. Optimal parameters of 
a duration of $\tau=164~\mu$s and amplitude $A=12.2~E_{\rm{r}}$ are found from both numerical simulation and empirical optimisation.

  Finally, after an additional time $T_2$, we repeat the initial splitting pulse, which now acts to recombine the wavepackets. The output of the interferometer is observed by recording an absorption image after 64~ms time-of-flight in a 15~G/cm magnetic levitation field to spatially separate the momentum states, and then determining the fractional populations of the $|0\hbar k\rangle$ and $|\pm2\hbar k\rangle$ momentum states by fitting independent Gaussians to the integrated image profiles. The population of the $|\pm4\hbar k\rangle$ is negligible and therefore is not included in the normalisation.

It is during the sequence of atom-optic pulses that we apply a magnetic field gradient by passing current through a coil comprising 5 turns of 1~mm diameter wire around a 2.4~cm diameter former, positioned with the coil coaxial with the interferometer splitting axis. The separation of the centre of the coil to the BEC was estimated to be 15$\pm$1~mm, with the uncertainty dominated by the difficulty of measuring the distance between the in-vacuo BEC and the ex-vacuo coil.

\section{\label{sec:coil}Coil model}

\begin{figure}
	\centering
	\subfloat{
		\includegraphics[clip,width=\columnwidth]{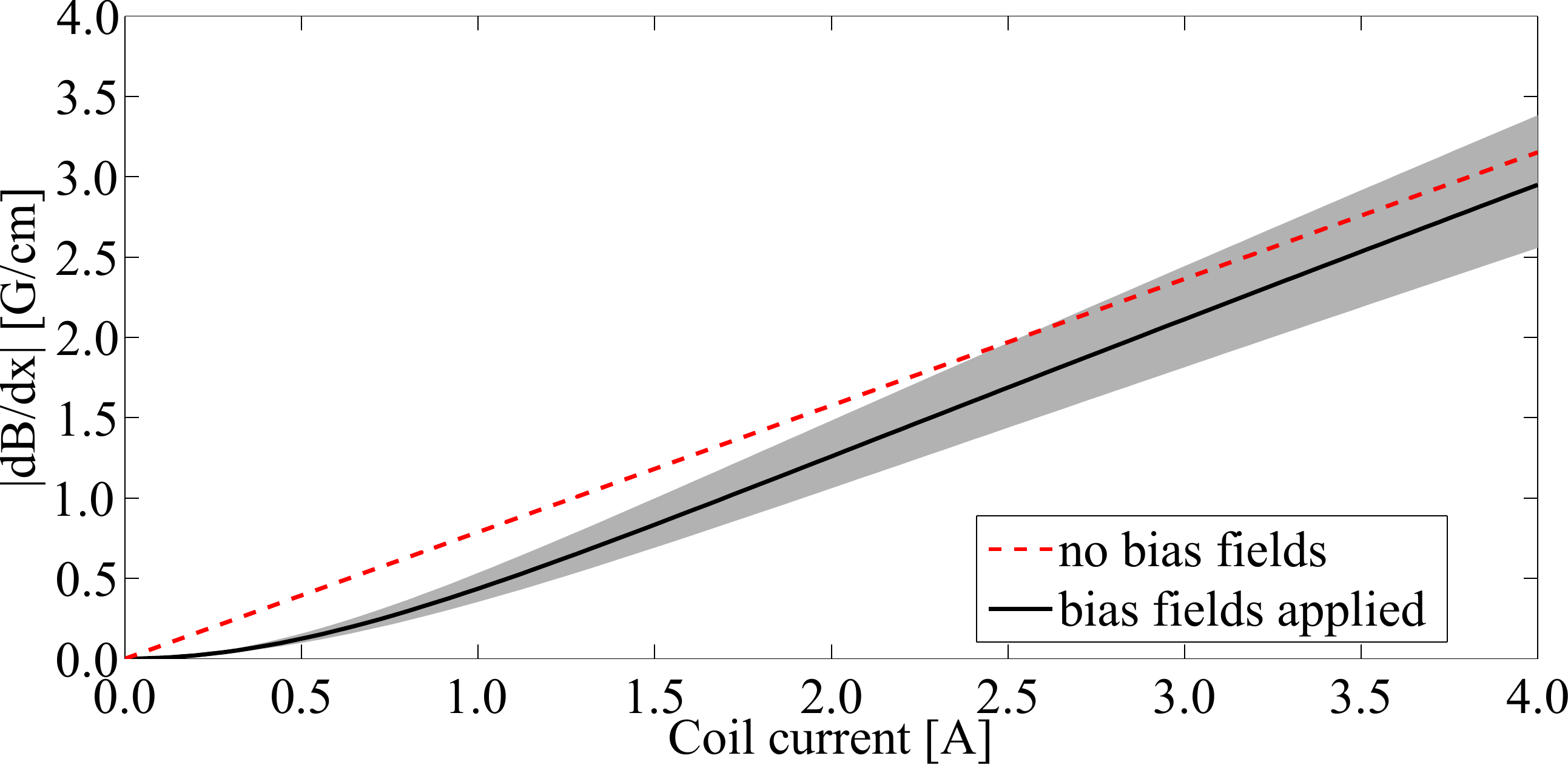}
		}
	\caption{Gradiometer coil model of magnetic field gradient magnitude, $|\partial B/\partial x|$, vs gradiometer coil current by application of the Biot-Savart law, with (solid black line) and without (dashed red line) applied bias field of 1.02~G along the z-axis. Shaded area indicates a coil position uncertainty of 1~mm.}
	\label{fig:model}
\end{figure}

We create a model of the effect of our gradient coil by application of the Biot-Savart law for an on-axis current loop. We can also account for the bias field applied during the experiment that is required to maintain an atomic quantisation axis, 1.02~G along the z-axis, which leads to the curvature of the solid black line in Fig.~\ref{fig:model}. Note that in the high current regime, by which we refer to currents above 1.5~A, the effect of the bias fields diminishes and the scaling between current, $I$, and magnetic field gradient, $\partial B/\partial x$, becomes approximately linear and we can write:
\begin{equation}
	\frac{\partial B}{\partial x} = \beta \left(I-0.52\right)~.
	\label{equ:_I_to_dBdx}
\end{equation}
From our model we predict $\beta$=0.85$\pm$0.11~G/(A$\cdot$cm) where the uncertainty is dominated by the uncertainty in estimating the position of the coil relative to the atoms.

\section{\label{sec:measurements}Pulsed mode operation}

The applied magnetic field gradient is varied by passing a current through the gradient coil. Here we operate the interferometer in pulsed mode in order to make an estimate of $\beta$. Whilst the application of the aforementioned bias fields means this scaling will be slightly non-linear at low current, as shown in Fig.~\ref{fig:model}, here we operate in the larger current regime ($I>1.5~$A) where the scaling maintains a linear relationship.

\begin{figure}
	\centering
	\includegraphics[clip,width=\columnwidth]{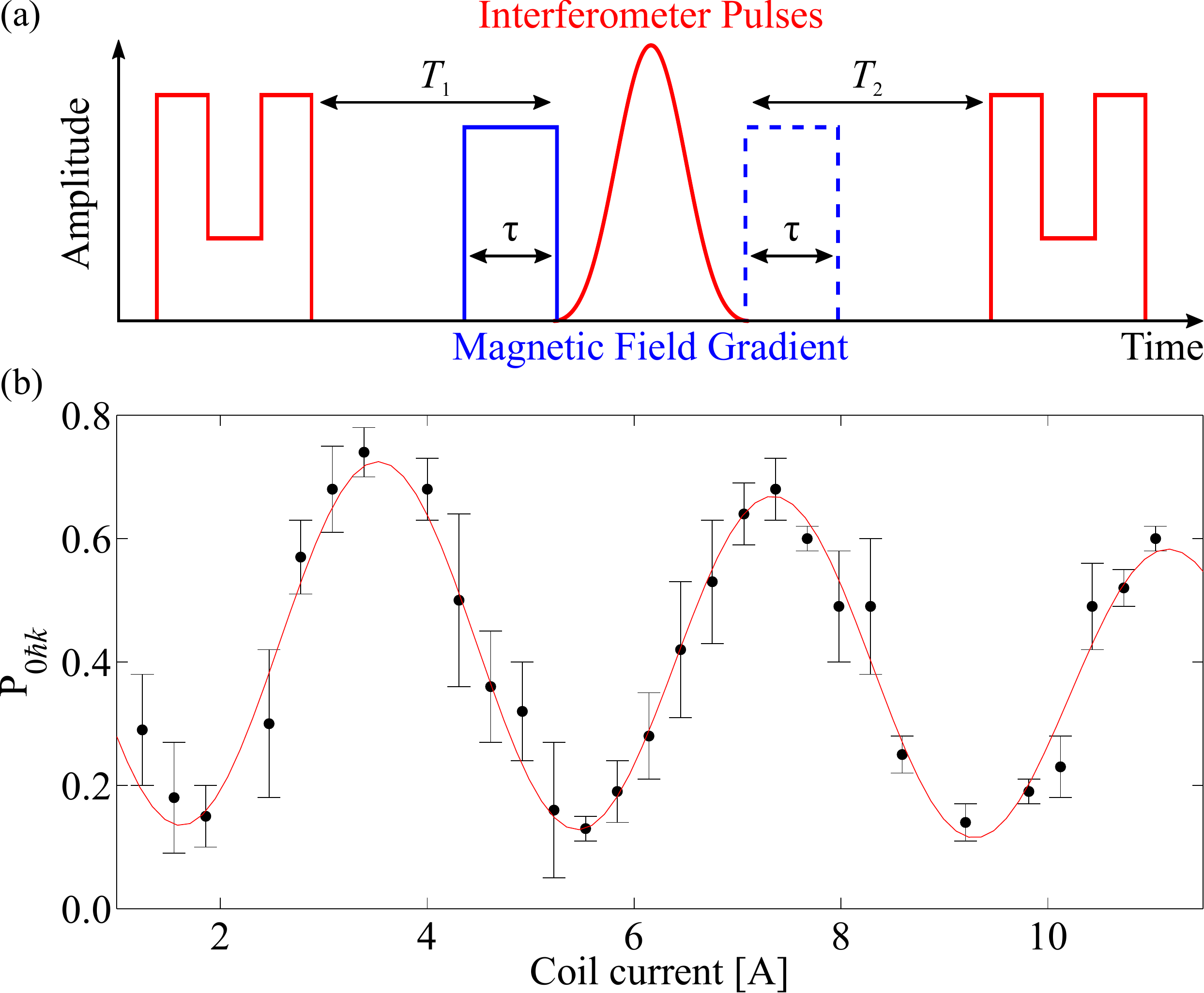}
	\caption{(a) A gradient pulse of varying amplitude is applied before or after the reflection and an interference fringe is observed. An example fringe for $T_1 = T_2 = 500~\mu$s with the gradient pulse applied after the reflection is shown in (b). Error bars indicate standard deviation}
	\label{fig:dBdx_pulses}
\end{figure}

Following the interferometer beam--splitter, see Fig.~\ref{fig:dBdx_pulses}a, we apply a gradient field of duration $\tau=260~\mu$s by passing a current through the gradient coil immediately preceding or following the reflection pulse, as indicated by the solid blue and dashed blue rectangle in Fig.~\ref{fig:dBdx_pulses}a, respectively. For a given interferometer duration the current through the coil is varied and the resultant fractional populations in the $|0\hbar k\rangle$ and $|\pm2\hbar k\rangle$ momentum states at the output of the interferometer determined. We label the $0\hbar k$ fractional population as $P_{0\hbar k}$ and we plot an example of this as a function of coil current in Fig.~\ref{fig:dBdx_pulses}, where $T_1=T_2=$~500~$\mu$s with the gradient applied before the reflection pulse.

The differential phase shift resulting from the temporally varying spatial separation, $\delta x\left(t\right)$, of the two momentum states in a magnetic field gradient can be written as~\cite{Wu05}:
\begin{equation}
	\Delta \phi_{\rm{mag}} = \int \frac{\mu~\frac{\partial |B|}{\partial x}~\delta x\left(t\right)}{\hbar}dt~,
	\label{equ:Wang_phase}
\end{equation}
with the magnetic moment of the test atom $\mu=m_{\rm{F}}g_{\rm{F}}\mu_{\rm{B}}$ where $m_{\rm{F}}$ is the magnetic Zeeman level of the atom, $g_{\rm{F}}$ is the Land\'{e} g-factor of the hyperfine atomic state, and $\mu_{\rm{B}}$ is the Bohr magneton. We use a simulation of the atomic trajectories to calculate the spatial separation $\delta x\left(t\right)$, then we numerically integrate Eq.~\ref{equ:Wang_phase}, with the scaling factor $\beta$ included in the calculation of $\partial B/\partial x$. This model then returns a phase, $\Delta\phi_{\rm{mag}}$, and we fit the following equation to the data:
\begin{equation}
	P_{0\hbar k} = A\cos\left(\Delta\phi_{\rm{mag}} + \phi_0\right)G\left(I\right)+P_0~,
	\label{equ:pulsed_model}
\end{equation}
where $A$ is the signal amplitude and $P_0$ is an amplitude offset. In our system the duration of the atom-optics pulses is significant with respect to the interferometer duration. As a result, the trajectories, and therefore the phase accumulation of the wavepackets during the atom-optic pulses, can be significant and non-trivial \cite{Griffin17,Antoine07}. To account for this we include a phase offset, $\phi_0$, in Eq.~\ref{equ:pulsed_model}. As these pulses have occurred in the absence of a gradient field, this phase offset is the same for all gradient coil currents. We also observe a decay in the visibility of the signal for increased current, and to account for this we include a Gaussian envelope, $G\left(I\right)$, centred around 0~A and with the standard deviation as a fit parameter. Loss of visibility at larger applied fields is attributed to acceleration effects increasing the velocity of the atomic wave-packets and reducing the efficiency of the interferometer recombination pulse. 

By performing interferometers with $T_1=T_2=$ 500~$\mu$s, 600~$\mu$s, and 700~$\mu$s, we determine $\beta$=0.79$\pm$0.01~G/(A$\cdot$cm) and $\beta$=0.81$\pm$0.01~G/(A$\cdot$cm) when the gradient is applied before and after the reflection pulse respectively. Both measured values of $\beta$ are in near agreement with each other, demonstrating the symmetry of the system, and are also consistent with the predicted value.

\section{\label{sec:continuous_mode_operation}Continuous mode operation}
\subsection{\label{subsec:theory}Theory}

\begin{figure}
	\centering
	\includegraphics[width=1\linewidth]{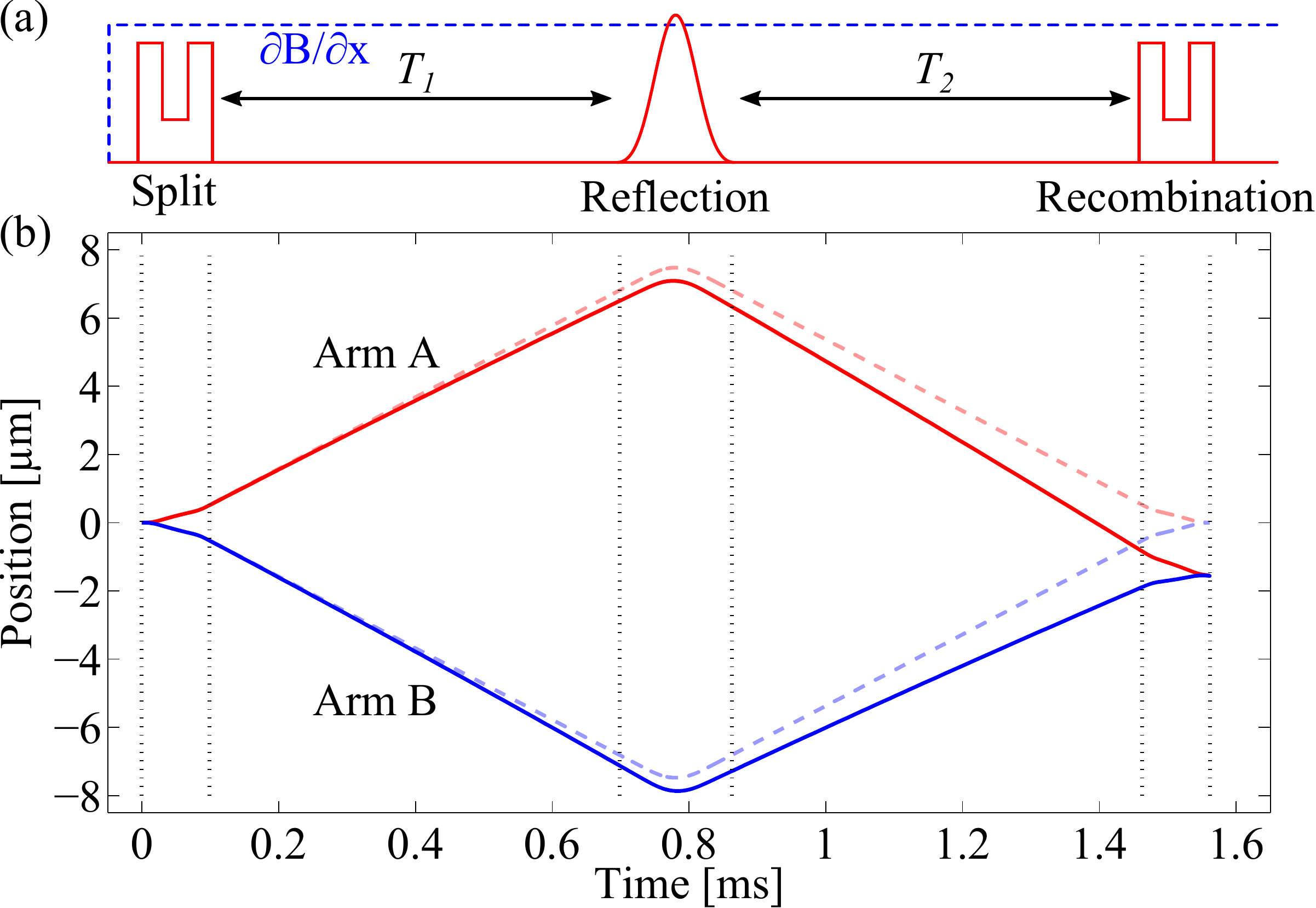}
	\caption{(a) Red solid line: Mach-Zehnder interferometer pulse sequence. Blue dashed line: applied magnetic gradient. (b) Simulation of atomic trajectory in the presence of 0~G/cm (dashed lines) and 2~G/cm (solid lines) static field gradient.}
	\label{fig:schematic}
\end{figure}

We now switch the interferometer to continuous mode operation where the field gradient is applied throughout the entire interferometry sequence as shown in Fig.~\ref{fig:schematic}a. The phase difference, $\Delta\phi$, between arms A and B (Fig. \ref{fig:schematic}b) at the output of the interferometer can be calculated by considering the classical action along each path $S_{\rm{cl}}^{\rm{A,B}}$:
\begin{equation}
	\Delta\phi = \frac{S_{\rm{cl}}^{\rm{A}} - S_{\rm{cl}}^{\rm{B}}}{\hbar}~,
	\label{equ:delta_phi}
\end{equation}
where
\begin{equation}
	S_{\rm{cl}}^{\rm{A,B}} = \int_0^{T} L^{\rm{A,B}}~dt,
	\label{equ:classical_action}
\end{equation}
with $T$ as the total interferometer duration, and $L$ is the Lagrangian which is given by the kinetic energy minus the potential energy ($E_{\rm{K}}-E_{\rm{P}}$)~\cite{Cohen-Tannoudji94}. For a symmetrical interferometer, such as the one reported here, $\Delta\phi=0$ because the phase accumulation due to the kinetic energy is equal to the phase accumulation due to the potential energy, i.e. $S^{\rm{A}}_{\rm{cl}}$ and $S^{\rm{B}}_{\rm{cl}}$ are equal~\cite{Cohen-Tannoudji94}. As a result the only phase detected in such a system is from the phase of the optical lattice produced by the splitting and recombination pulses. The positional shift caused by the gradient field results in the atoms being in a different optical potential at the point of recombination relative to the optical potential of the splitting pulse, as indicated in Fig. \ref{fig:schematic}. It is therefore reasonable to consider the optical lattice as a `ruler' against which the wavepacket centre-of-mass is measured.

The centre-of-mass displacement of the BEC can be written as
\begin{equation}
	s=\frac{\Delta\phi}{k_{\rm{eff}}}~,
	\label{equ:output_phase}
\end{equation}
where $\phi$ is the phase of the interferometer output and $k_{\rm{eff}}=4\pi\cos(\theta)/\lambda$, where $\lambda$ is the wavelength of the interferometry laser and $\theta$ is the beam angle relative to the interferometry axis (see Fig.~\ref{fig:experiment_diagram}). Assuming zero initial velocity, the displacement can also be written as $s=aT^2/2$. We can therefore write
\begin{equation}
	\Delta\phi = \frac{1}{2}a\,k_{\rm{eff}}T^2~,
	\label{equ:main_equ}
\end{equation}
with the acceleration given by
\begin{equation}
	a = -\frac{\mu}{m}\frac{\partial B}{\partial x}~,
	\label{equ:acceleration}
\end{equation}
where $m$ is the atomic mass and $T$ is the interferometer duration.

A simulation of the atomic trajectory is shown in Fig.~\ref{fig:schematic}b with $\partial B/\partial x$=2~G/cm. We also include the trajectories during the atom-optics pulses, which are determined by considering the average centre-of-mass motion from simulations~\cite{Griffin17}.

\subsection{\label{subsec:t_squared}Experiment}

Here we apply the gradient for the entire interferometer duration, where, due to the finite switching time of the gradient coil, we turn it on 50~$\mu$s before the start of the splitting pulse, which will result in an initial velocity. For long interferometer durations such that $T\gg t$, the phase accumulation of the interferometer output will be linear with $T^2$, as predicted by Eq.~\ref{equ:main_equ}. Therefore, we can write
\begin{equation}
 	P_{0\hbar k}=A\cos\left(k_{\rm{eff}}\frac{1}{2}aT^2 + \phi_0\right)G\left(T\right)+P_0~,
 	\label{equ:simple_fit}
\end{equation}
where $A$ is amplitude, and $P_0$ is an arbitrary amplitude offset. As above, we also observe decay in the visibility of the signal with time, and in order to obtain a good fit to the data we include a Gaussian envelope $G\left(T\right)$ centred around $T=0$ with a standard deviation of 2~ms. A phase offset, $\phi_0$, is also included in the model to account for non-trivial phase evolution during the atom-optic pulses~\cite{Griffin17,Antoine07}.

\begin{figure}
	\centering
	\includegraphics[clip,width=\columnwidth]{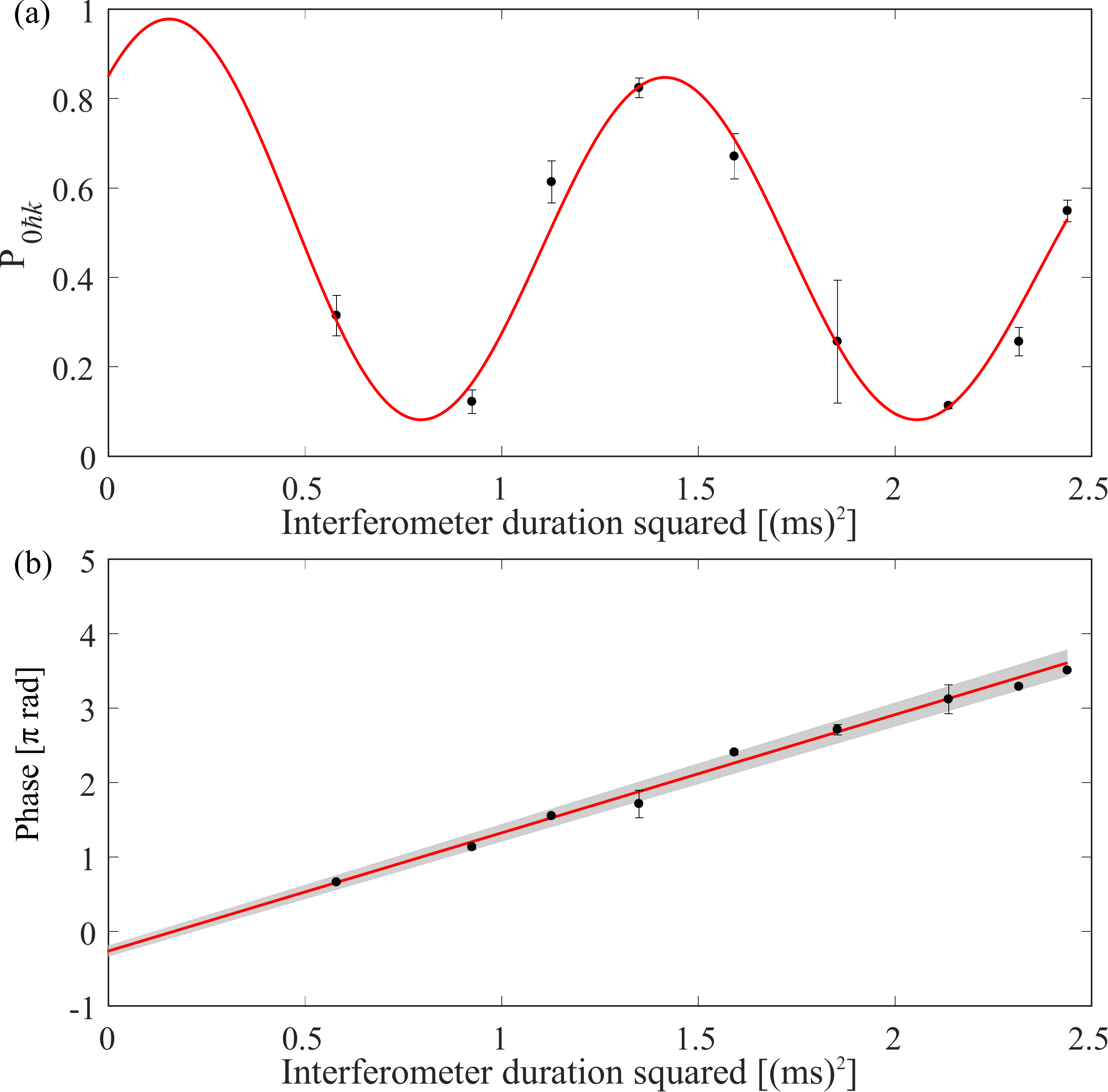}
	\caption{(a) Gradiometer operating in continuous mode whilst applying a coil current of 1.9~A, with Eq.~\ref{equ:main_equ} fitted to the data. (b) We convert population, $P_{0\hbar k}$, to phase and fit with a linear model. The error bars indicate standard error of each data point (where not visible these are included within the data point) while the shaded area around the straight line represents the standard deviation of the fit as a result of the uncertainty in the original fitted model. Note that the uncertainty in the phase of the data is greatly increased where the gradient of the fitted model is close to zero and for increased interferometer durations.}
	\label{fig:t_squared}
\end{figure}

To test this we perform a temporally symmetric interferometer sequence of varying duration whilst applying a fixed gradient coil current and determine $P_{0\hbar k}$. We plot these data as a function of interferometer duration squared in Fig.~\ref{fig:t_squared}a and fit Eq.~\ref{equ:simple_fit}. The $P_{0\hbar k}$ data are then converted to phase and shown in Fig.~\ref{fig:t_squared}b, where the straight line is the linearized version of Eq.~\ref{equ:simple_fit}.

\subsection{\label{subsec:residual}Determination of residual forces}

In any apparatus designed for precision measurement it is important to characterize the system to determine any residual/unintentional forces. Here, we present a method of measuring such forces using the atoms themselves. In our system the residual forces, and therefore accelerations, are small such that an unbiased interferometer would be insensitive to them; the signal frequency would be too low to accurately measure. Therefore, we bias our interferometer such that we can determine the magnitude and direction of a stray magnetic field gradient as well as an inertial acceleration.

It should be noted that the sign of the phase shift due to magnetic fields, Eq.~\ref{equ:Wang_phase}, is independent of the vector component of the magnetic field. This can be understood by recognising that the atomic state used, $|F=2, m_F=2\rangle$ is a weak-field seeker, which will always be repelled from the coil, irrespective of the direction of current flow. This is in contrast to other inertial forces, such as gravity, that have an effect on the sign of the phase shift
\begin{equation}
	\Delta \phi_{\rm{grav}} = \int \frac{m~ g_x\,\delta x\left(t\right) }{\hbar}dt~,
	\label{equ:gravitational_phase}
\end{equation}
where the term $g_x$ is the projection of the local acceleration due to gravity, $g$, onto the interferometry axis; i.e., $g_x=g\cos(\gamma)$, where $\gamma$ is the angle between the x-axis and the gravitational acceleration.

In section~\ref{subsec:t_squared} we ignored the effect of the initial velocity of the atoms as a result of applying the gradient field $t=$~50~$\mu$s before the start of the splitting pulse.  The rise time of the coil current has an exponential form of $1-\exp\left(-t/\tau\right)$ with $\tau=5.7~\mu$s, and therefore $u=a\,t$ becomes $u=a\left(t+\tau\exp\left(-\tau/t\right)-\tau\right)$. This results in an effective shortening of $t$ from 50~$\mu$s to 44~$\mu$s. When we include this initial velocity, the phase accumulation of the interferometer goes as $k_{\rm{eff}}\left(uT + \frac{1}{2}aT^2\right)$, and $P_{0\hbar k}$ will have the form
\begin{equation}
	\resizebox{.88\hsize}{!}{$ 
	P_{0\hbar k} = A\cos\left(k_{\rm{eff}}\left(uT + \frac{1}{2}aT^2\right)+\phi_0\right)\\ G\left(T\right) + P_0~.
	$}
	\label{equ:fit_function}
\end{equation}

\begin{figure}
	\centering
	\subfloat{
		\includegraphics[clip,width=\columnwidth]{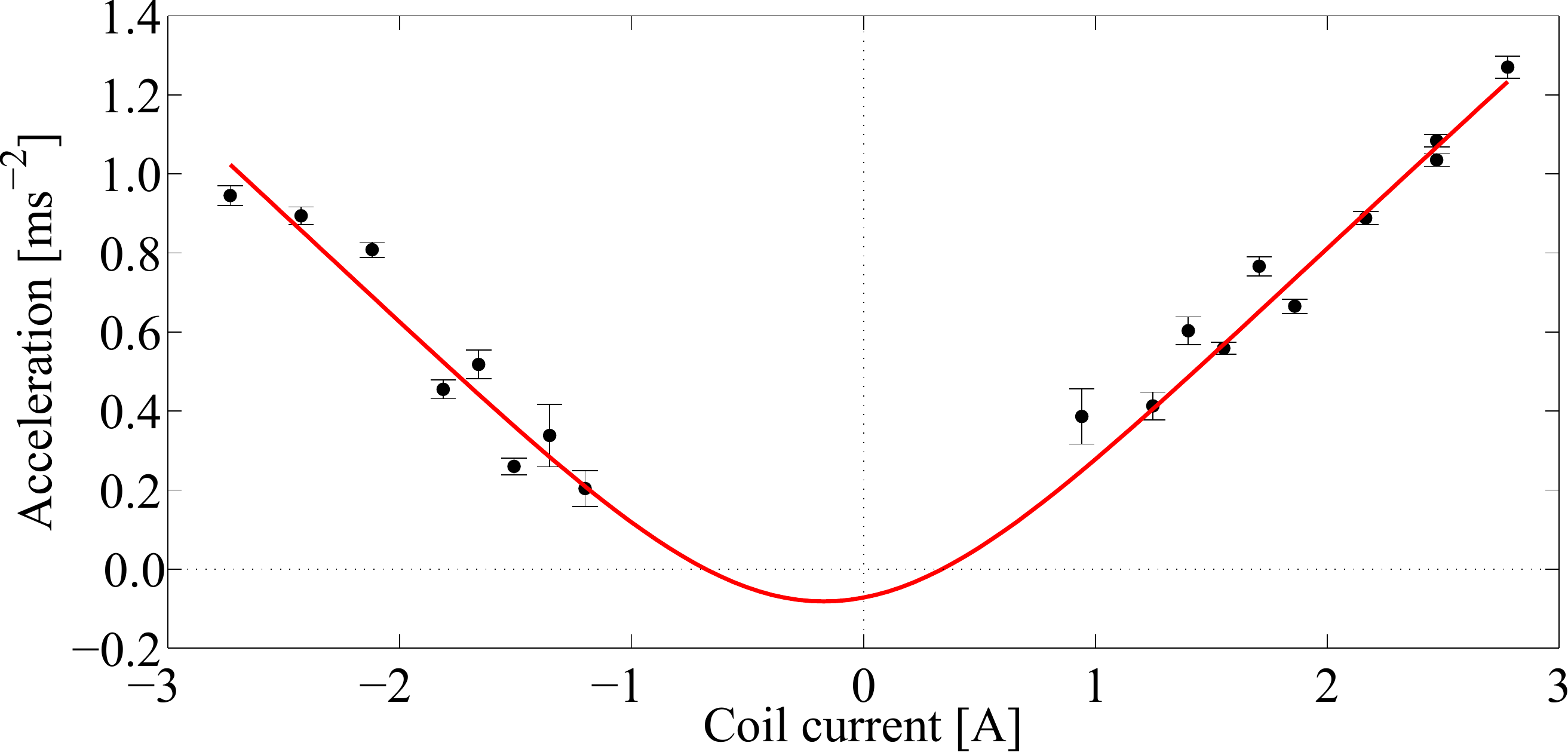}
		}
	\caption{Measured acceleration for varying gradient coil currents. Error bars indicate standard deviation. The fitted curve is a model of our gradient coil with coil-atom distance, acceleration offset, and current offset as free parameters. We determine the effective coil-atom distance to be 15.0$\pm$0.2~mm.}
	\label{fig:residual}
\end{figure}

For a range of gradiometer coil currents we probe our system for varying interferometer durations, $T$. To these data we fit Eq.~\ref{equ:fit_function} and extract an acceleration and plot the data in Fig.~\ref{fig:residual}. We fit our coil model to these data with the coil distance, acceleration offset and current offset as the free parameters. The coil distance is found to be 15.0$\pm$0.2~mm from the fit, in good agreement with the spatial measurement.

The residual magnetic field gradient is determined entirely by the current offset (0.17$\pm$0.02~A)
of the fitted curve in Fig.~\ref{fig:residual}, which is the current that cancels out the residual field. From our coil model, along with the fitted coil-atom distance we calculate the value of this field gradient to be 15$\pm$2~mG/cm.

Residual forces that are non-magnetic (gravitational, electrostatic, etc.) can be determined entirely from the acceleration offset of the fitted model, where a negative offset indicates a force toward the coil and vice-versa. From the acceleration offset in the fitted model we determine a residual acceleration of -0.08$\pm$0.02~m/s$^{2}$. If we attribute this to a tilt in the interferometer axis with respect to gravity, it is equivalent to an angle of $\gamma=-0.5\pm$0.1\si{\degree}. The negative sign indicates that the angle of the interferometer axis with respect to gravity is such that, in the absence of an applied field, the atoms would accelerate towards the gradient coil. Without the use of this method it would be not be possible to characterize the residual forces with our interferometer in its current, horizontal implementation; in order to make a direct measurement of our residual non-magnetic force we would require interferometer durations of $\sim$6~ms to observe just one complete fringe, at which time the atoms would have fallen out of the interrogation region.

For these data we are also able to extract a $\beta$ value, and we find $\beta=0.81\pm0.02$~G/(A$\cdot$cm). This is consistent with our initial estimate of $\beta$=0.85$\pm$0.11~G/(A$\cdot$cm) and agrees well with our measured values of $\beta$=0.79$\pm$0.01~G/(A$\cdot$cm) and $\beta$=0.81$\pm$0.01~G/(A$\cdot$cm) from operating the interferometer in pulsed mode in section~\ref{sec:measurements}.

\section{\label{sec:conclusions}Conclusions}

In summary we have constructed a matter-wave interferometer and have demonstrated its ability to make measurements of magnetic field gradients. Two modes of operation have been demonstrated: pulsed mode and continuous mode. Both methods show good agreement in measuring the scaling factor, $\beta$, relating the gradient coil current to magnetic field gradient, and these values are also consistent with our initial estimate. In addition, we have created a method for measuring low ambient fields that would be impossible to measure using an unbiased cosine-sensitive interferometer as the required interferometer durations would be too long. One can therefore envisage the measurement of even smaller ambient fields by extending the available interrogation time. In our case this would require the use of interferometry beams with a larger waist, or the use of a magnetic levitation field, which would potentially couple noise into the system. The techniques presented provide excellent tools for the characterisation of applied magnetic field gradients and residual accelerating fields, crucial in precision measurement.

\section{\label{sec:conclusions}Acknowledgements}
This work was funded by the UK National Quantum Technology Hub in Sensing and Metrology, through the EPSRC (EP/M013294/1).


%

\end{document}